\documentclass[useAMS,twocolumn,usenatbib]{mnras}

\usepackage{graphicx,enumitem,color,hyperref}
\usepackage{xcolor}
\usepackage{amssymb,amsmath,xtab}

\newcommand{\ha}{H$\alpha$}
\newcommand{\sbline}{$\rm erg~s^{-1}~cm^{-2}~arcsec^{-2}$}

\def\HI{\hbox{H~$\scriptstyle\rm I\ $}}

\def\nHI{{\rm HI}}

\def\spose#1{\hbox to 0pt{#1\hss}}
\def\lta{\mathrel{\spose{\lower 3pt\hbox{$\mathchar"218$}}
     \raise 2.0pt\hbox{$\mathchar"13C$}}}
\def\gta{\mathrel{\spose{\lower 3pt\hbox{$\mathchar"218$}}
     \raise 2.0pt\hbox{$\mathchar"13E$}}}

\hypersetup{draft} 
\begin{document}
\title[UV background at $z=0$]{MCMC determination of the cosmic UV background at $z\simeq 0$ from \ha\ fluorescence.}
\author[D. Caruso et al.] {Davide Caruso$^{1}$, Francesco Haardt$^{1,2}$, Michele Fumagalli$^{3,4}$,  Sebastiano Cantalupo$^5$ \vspace{2mm}
\\$^1$DiSAT, Universit\`a dell'Insubria, Via Valleggio 11, I-22100 Como, Italy
\\$^2$INFN, Sezione di Milano-Bicocca, Piazza delle Scienze 3, I-20123 Milano, Italy
\\$^3$Institute for Computational Cosmology, Durham University, South Road, Durham DH1 3LE, UK 
\\$^4$Centre for Extragalactic Astronomy, Durham University, South Road, Durham DH1 3LE, UK
\\$^5$Department of Physics, ETH Zurich, Wolfgang Pauli Strasse 27, CH-8093 Zurich, Switzerland}
\date{\today} 
\maketitle

\begin{abstract} 
In a recent paper \citep{fumagalli17} we reported on the detection of a diffuse H$\alpha$ glow in the outskirts of the nearby, edge-on disc galaxy UGC 7321 observed with the Multi Unit Spectroscopic Explorer (MUSE) at the ESO Very Large Telescope. By interpreting the H$\alpha$ emission as fluorescence arising from hydrogen ionised by an external (i.e., extragalactic) radiation field, we estimated the UV background (UVB) intensity in terms of \HI ionisation rate (per ion) at $z\simeq 0$ to be in the range $\Gamma_\nHI\sim 6-8\times 10^{-14}$ s$^{-1}$. In the present work, by performing radiative transfer calculations over a large set of models of the gaseous disc of UGC 7321, we refine our estimate and through an MCMC analysis derive a value for the photoionisaton rate of $\Gamma_\nHI=7.27^{+2.93}_{-2.90}\times 10^{-14}$ s$^{-1}$. In particular, our analysis demonstrates that this value is robust against large variations in the galaxy model and that the uncertainties are mainly driven by the errors associated with the observed \ha\ surface brightness. Our measurement is consistent with several recent determinations of the same quantity by a completely independent technique (i.e., flux decrement analysis of the Ly$\alpha$-forest), and support the notion that the low redshift UVB is largely dominated by active galactic nuclei (AGNs), possibly with no need of further contribution from star forming galaxies.   
\end{abstract}


\section{Introduction}
\label{sec:introduction}
The reionisation of the all-pervading intergalactic medium (IGM), the repository of most of the baryons across the history of the Universe, is a landmark event in the cosmic history of structure formation. Modern observations of the IGM have provided several tests of the $\Lambda$CDM paradigm, including a measurement of the power spectrum, upper limits to the neutrino masses, and a measure of baryonic acoustic oscillations \citep[e.g.,][]{mcdonald05,viel10,slosar13}. 

Most of our understanding of IGM physics, and its implication for galaxy formation and metal enrichment, depends critically on the properties of the cosmic ionising UV background (UVB), the integrated UV emission from all possible emitting sources in the Universe. Massive stars in young star forming galaxies and accreting supermassive black holes in active galactic nuclei (AGNs) are the most obvious sources of ionising UV radiation \citep[e.g.,][]{miralda90,HM96}, still their relative importance across the cosmic time is not firmly established \citep[see, e.g.,][and references therein]{kulkarni18}. 

Recent observational and theoretical progress are forging a coherent description of the thermal state and ionisation degree of the IGM. 
The UVB reionised the hydrogen component of the IGM by $z\simeq 6$ \citep[e.g.,][]{davies18,planck18}, while the double reionisation of helium occurred later, at $z\gta 3$ \citep[e.g.,][]{worseck16}, because of the reduced cross section and higher ionisation potential. In the post reionisation Universe, the UVB keeps the bulk of the IGM ionised \citep[e.g.,][]{gunn65,bolton07}, regulates its temperature \citep[e.g.,][]{theuns02}, and sets a characteristic mass below which haloes fail to form galaxies \citep[e.g.][]{okamoto08}. 

In the absence of firm observational constraints on $\Gamma_\nHI$ the current parametrisation of the UVB relies mostly on 1D radiative transfer calculations that follow the build-up of the UVB accounting for sources and sinks of radiation. These models have input parameters that are difficult to measure, such as the emissivity and escape fraction of ionising photons from galaxies and AGNs, and the distribution of \HI\ absorbers \citep[e.g.,][]{HM96,shull99,faucher09,HM12}. Therefore, different models predict values of $\Gamma_{\rm HI}$ that differ by factors of a few, primarily because the observational data that enter the modelling are not well known. 
As an example, the most recent models of the low-redshift UVB \citep{khaire15,MH15,onorbe17,puchwein18}, adopting an updated AGN emissivity predict a value of $\Gamma_\nHI$ at $z\simeq 0$ which lies a factor $\simeq 2$ above the value predicted by \citet{HM12}. However, independently upon details, all recently proposed models agree with 
a UVB which is largely dominated by AGNs at low resdhifts.

The amplitude of the UVB can be measured by three different methods. Firstly, the hydrogen ionsation rate per ion, $\Gamma_\nHI$, can be inferred using the so called "proximity effect" \citep[e.g.,][]{murdoch86,bajtlik88}, i.e., by determining out to which distance the local ionisation front of a single QSO outshines the UVB. Secondly, constraints on $\Gamma_\nHI$ can be derived from the statistical comparison of the observed Ly$\alpha$ forest to numerical simulation predictions \cite[e.g.,][]{rauch97,calverley11}. This method, 
albeit affected by systematic uncertainties, offers the primary constraints on $\Gamma_\nHI$ both at low 
\citep[e.g.,][for $z\lta 1$]{khaire15,shull15,viel17,gaikwad17} and high \citep[e.g.,][for $2\lta z \lta 6$]{bolton05,kirkman05,faucher08a,wyithe11,becker13,davies18} redshifts. Note however that observing the Ly$\alpha$ forest at $z\lta 1$ requires UV spectroscopy from space.

A third method, the detection of fluorescence Ly$\alpha$ from the recombining IGM overdensities in ionisation equilibrium with the UVB, proved to be extremely challenging because of the very low surface brightness (SB) involved \citep[see, e.g.,][]{gould96,cantalupo05,rauch08,gallego18}. However, fluorescence could also be detected in H$\alpha$ in the local Universe, hence without the redshift SB dimming effect as in the case of Ly$\alpha$, by observing the ionisation front in clouds photoionised by the UVB \citep[][]{vogel95,donahue95,weymann01}, or in the outskirts of the \HI discs of galaxies \citep[e.g.][]{maloney93,dove94,bland97,cirkovic99,madsen01,bland17}. Using this technique, \citet{adams11} targeted the nearby edge-on galaxy UGC 7321, obtaining an upper limit for $\Gamma_\nHI$.

In a recent paper \citep{fumagalli17} we described the results from a pilot MUSE \citep{bacon10} observation at the VLT, following the line proposed by \citet{adams11}, in the attempt to measure the UVB intensity by searching for the \ha\ recombination line at the edge of the \HI\ disk in the nearby, edge-on disc galaxy UGC 7321. We described how an emission line was detected in a deep 5.7-hour exposure  at $\lambda \simeq 6574~$\AA,
which is the wavelength where H$\alpha$ is expected given the \HI\ radial velocity of UGC 7321. Despite the presence of a skyline at similar wavelengths, we consistently recovered the \ha\ signal within
data cubes reduced with different pipelines, and within data cubes containing two independent sets of exposures. We concluded that we indeed detected \ha\ recombination radiation at a level of $(1.2 \pm 0.5) \times 10^{-19}~$\sbline. Assuming photoionisation from the the UVB as origin of the observed signal, through 1-D radiative transfer calculations and the joint analysis of spatially-resolved
\HI\ column density and \ha\ SB maps, we translated the observed SB
into a value for $\Gamma_\nHI$ of order of  $(6-8)\times 10^{-14}~\rm s^{-1}$, consistent with the values inferred from the statistics of the low-redshift Ly$\alpha$ forest. 

In this paper we present a more refined statistical analysis of our data based on a Monte Carlo Markov's Chain (MCMC) procedure, leading to a more precise and robust determination of the UVB in terms of $\Gamma_\nHI$ and its associated error.

\section{Data production and analysis}
A thorough discussion of our H$\alpha$ fluorescence line detection, with all relevant technical information, can be found in \citet{fumagalli17}. Here we simply summarise the main key points.

The MUSE observations of the edge-on disk galaxy UGC 7321 have been acquired between 2015 June and 2016 January at the UT4 VLT, as part of the programme ID 095.A-0090 (P.I. Fumagalli). The location of the MUSE pointing was chosen to overlap with the region where the H$\alpha$ SB was expected to be maximised by the limb brightening  according to the galaxy model presented in \citet{adams11}, although our analysis subsequently revealed an offset. 

After standard data reduction of the exposures, we applied and compared three different pipelines in order to obtain our final data cube: i) the ESO MUSE pipeline \citep[v1.6.2;][]{weilbacher14}; ii) the cubextractor package (Cantalupo 2018, in preparation), following \citet{fumagalli16} and \citet{borisova16}, and iii), a custom post-processing pipeline combined to the Zurich Atmosphere Purge (ZAP) package \citep{soto16}. Moreover, to fully assess the performance of the data reduction techniques employed, we made use of mock data cubes that included emission lines injection at desired wavelengths followed by line recovery procedures. 

\ha\ recombination signal was then searched both in 2-D maps and in the mean spectra constructed by averaging the flux from all the pixels inside a specified region. Further tests on the origin of the detected signal were performed. Based on this analysis we concluded that we detected an emission feature at $\lambda \sim 6574~$\AA\ consistently in two independent data reductions and in two independent sets of exposures with different
instrument rotations. The emission overlaps with the location where \HI\ is detected with $N_{\rm HI} \gtrsim 10^{19}~\rm cm^{-2}$. 
Altogether, these pieces of evidence corroborated the detection in the outskirts of UGC 7321 of
an extended low SB signal consistent with \ha\ recombination radiation of gas in photionisation equilibrium with the cosmic UVB.

\section{Galaxy models}
In order to translate the \ha\ signal into a value of $\Gamma_\nHI$, a model for the hydrogen disc of our target galaxy is needed. As in \citet{fumagalli17}, and following \citet{adams11}, we adopt a simple, double exponential disc, of the form  
\begin{equation}
n_{\rm H}(R,z)=n_{\rm H,0}\exp{(-R/h_R)}\exp{(-|z|/h_z)},
\end{equation}
where $n_{\rm H}(R,z)$ is the total hydrogen number density in cylindrical coordinates ($R,z$), $n_{\rm H,0}$
defines the central density, while $h_R$ and $h_z$ are, respectively, the radial scale-length and
vertical scale-height of the disc. The galaxy disc, inclusive of an He component at cosmic fraction, is then irradiated by an external ionising radiation field, and the ionisation and temperature structures are solved under the assumption of ionisation and thermal equilibrium. 
   
  In order to ease the calculation, we solve for the vertical ionisation and
temperature of the disk at a fixed radial distance $R$ assuming a two-sided plane parallel geometry, hence effectively reducing the 3-D radiative transfer problem to a series of much simpler 1-D calculations. The final full structure of the gaseous disk in terms of temperature and ion fractions is thus reconstructed combining results of calculations with plane-parallel geometries at different $R$ \citep[details of the adopted radiative transfer (RT) scheme are described in][]{HM12}. 
The ionisation and thermal vertical structures are solved iteratively for an input
power-law spectrum with spectral slope $1.8$. Ionisation equilibrium is achieved by balancing
radiative recombinations with photoionisation, including the formation and propagation of recombination radiation from
\ion{H}{II}, \ion{He}{II} and \ion{He}{III}. For the thermal structure, photo-heating is balanced by free-free, collisional
ionisation and excitation, and recombinations from \ion{H}{II}, \ion{He}{II}, and \ion{He}{III}. 
Our RT model makes a series of simplifying assumptions, e.g., ignoring metals and clumpiness. However, as our analysis will show, $\Gamma_\nHI$ is primarily constrained by the SB at the ionisation front, which does not appear to be very sensitive to the detail of the RT calculation, as shown also in \citet{bland17}.

The \ha\ volume emissivity is then calculated as 
\begin{equation}
\epsilon_{\rm{H}\alpha}(R,z)=h\nu_{\rm{H}\alpha} \alpha^{\rm eff}_{\rm{H}\alpha}(T)\, n_{\rm p}(R,z)n_{\rm e}(R,z)\:,
\end{equation}
where $n_{\rm p}$ and $n_{\rm e}$ are the proton and electron number densities,
and $\alpha^{\rm eff}_{\rm{H}\alpha}$ is the \ha\ effective case A recombination coefficient \citep{pequignot91}\footnote{Assuming instead case B (i.e., all ion transitions to the ground state are optically thick) results in $\alpha^{\rm eff}_{\rm{H}\alpha}$ which is a factor $\simeq 1.5$ larger for the typical temperature ($\simeq 10^4$ K) we find at the I-front. This would imply a similar {\it reduction} in the estimated value of $\Gamma_\nHI$.}.
The \HI\ column density and the \ha\ SB maps are then derived by 
integrating the neutral hydrogen density $n_\nHI$ and $\epsilon_{\rm{H}\alpha}$ along the line-of-sight, assuming an inclination angle of 83$^o$ \citep{adams11}. Details concerning the relations that connect cylindrical coordinates $(R,z)$ to the projected position  $(b_1,b_2)$ can be found in \citet{fumagalli17}.

We run a total of 7,056 galaxy models, varying $n_{\rm H,0}$ from 0.1 to 6.0 cm$^{-3}$ in sixteen unevenly spaced intervals, the disk scale-length in the interval $h_R = 1.3-2.9~\rm kpc$ in nine steps of $200~\rm pc$,
and the disk scale-height in the interval $h_z = 100-700~\rm pc$ in seven steps of $100~\rm pc$. For each combination of disk parameters, we perform the radiative transfer calculation for seven different
values of the UVB intensity, $\Gamma_{-14} = (1,2,4,6,8,12,16)$, where $\Gamma_{-14}\equiv \Gamma_\nHI /(10^{-14}~\rm s^{-1})$.

\section{Results and Discussion} 

The inclusion of 21-cm data is instrumental in breaking the degeneracy between the \ha\ SB and the hydrogen density. Indeed, if taken separately, \HI and \ha\ data can  provide only weak 
constraints on $\Gamma_\nHI$. In principle, MUSE could provide spatially resolved SB maps, and one could derive tight constraints on $\Gamma_\nHI$ through a joint analysis of HI and H$\alpha$ 2-D maps.
In practice, the large uncertainty of current line measurements makes a detailed 2-D analysis unnecessary at this stage. We therefore make use of the integrated H$\alpha$ flux in MUSE field of view (FOV). Specifically, we integrated the 2-D H$\alpha$ flux within the region defined by the 21-cm contour at $N_\nHI=10^{19}$ cm$^{-2}$ (the reference "I-front" contour), obtaining a SB of $(1.2\pm 0.5)\times 10^{-19}$ \sbline. The associated error is dominated by systematics in proximity to a bright sky line at $\lambda \simeq 6577$ \AA\ and at the edges of the FOV. We require that the \ha\ SB measured in models in the same area matches the observed value within the errors.   

\begin{figure}
	\centerline{\includegraphics[width=0.5\textwidth]{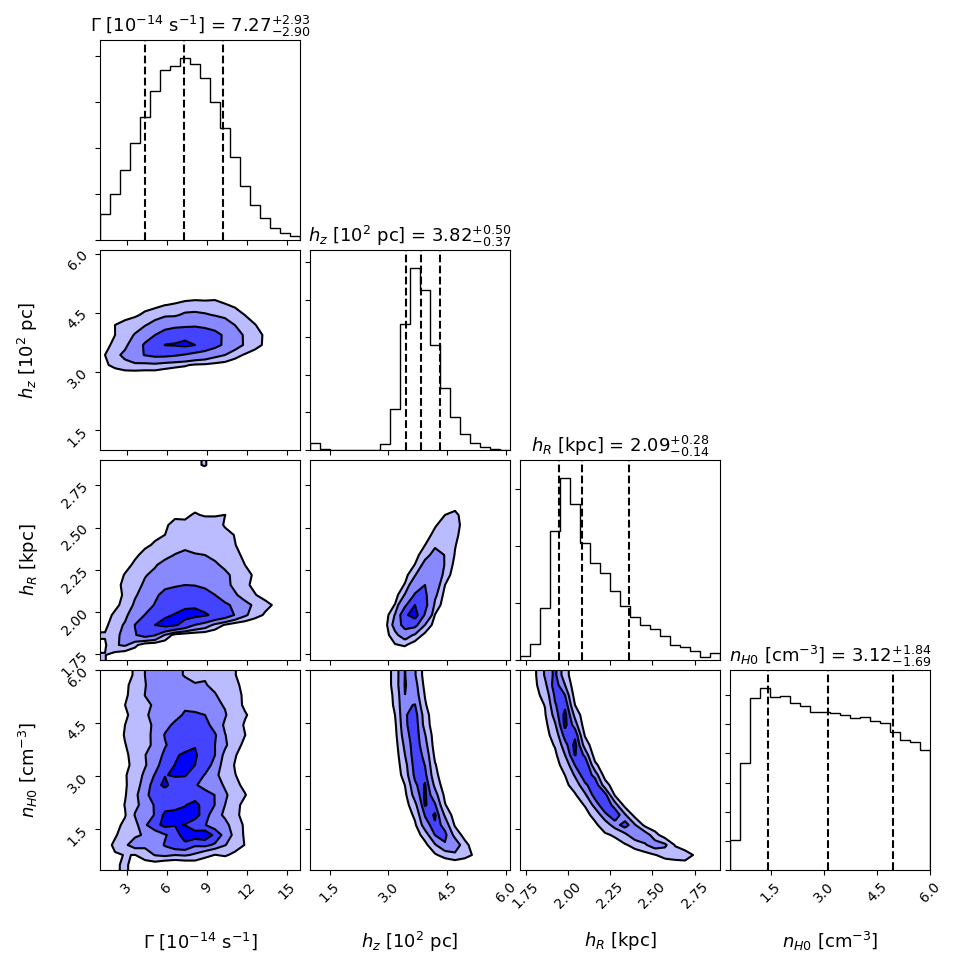}}
	\caption{Corner plot from the MCMC statistical analysis of our \ha\ detection plus \nHI\ contours. Systematics on the measured SB reflects into the relatively large uncertainties of the UVB estimate. However $\Gamma_\nHI$ is largely independent upon galaxy model details. Note the degeneracy of the parameters defining the galaxy gas distribution, in particular $n_{\rm H,0}$ and $h_R$.}
	\label{fig:corner}
\end{figure}

Concerning $N_\nHI$ data, we do not directly employ a statistical analysis of 21 cm maps. Rather, we identify three different parameters that, jointly, constrain the position of the ionisation front. Firstly, we demand that the semi-major and semi-minor axes measured in models at the I-front contour match the observed values of $b_{1,{\rm HI}} = 12.4 \pm 0.1~\rm kpc$ and $b_{2,{\rm HI}} = 2.3 \pm 0.2~\rm kpc$ within the associated errors. Secondly, we ask that the area in the MUSE FOV within the same contour level measured in models matches (in terms of total number of MUSE pixels) the observed value of $15,000\pm 150$ pixels. The associated error of 1\% is meant to account for an uncertainty of $\pm 1$ pixel around the position of the I-front contour.
 
The posterior distributions of the four model parameters, i.e., $n_{\rm H, 0}$, $h_R$, $h_z$ and $\Gamma_\nHI$, are then derived through an MCMC analysis\footnote{We use the {\texttt{emcee}} code by \citet{foreman13}.}, assuming flat priors. In order to speed-up the MCMC analysis, we construct multidimensional functions that linearly interpolate in the 4-D model parameter space, rather than performing the calculation at each step. Results in terms of the so-called corner plot are shown in Fig.~\ref{fig:corner}. The value $\Gamma_{-14}=7.27^{+2.93}_{-2.90}$ is consistent with what we derive in \citet{fumagalli17}, hence higher than the upper limit reported by \citet{adams11}. The UVB intensity we derive is consistent with estimates of $\Gamma_\nHI$ inferred from flux decrement analysis of the low-redshift ($z\lta 0.4$) Ly$\alpha$ forest in the absorption spectra of distant quasars, specifically \citet{shull15}, \citet{viel17}, and \citet{gaikwad17}. The relatively large error on our UVB estimate reflect the uncertainties, mainly due to systematics, in our measurement of the \ha\ SB.


Recent models of the UVB \citep[e.g.,][]{khaire15,MH15,onorbe17,puchwein18}, adopting an updated, larger AGN emissivity at low-redshift compared to \citet{HM12}, predict a value of $\Gamma_\nHI$ consistent with our estimate. In such models the UVB is largely dominated by AGNs up to $z\lta 2.5$, with star forming galaxies giving a negligible contribution.  

Fig.~\ref{fig:corner} shows that the central hydrogen density is degenerate with $h_R$ and $h_z$, as different combinations of the three parameters result in the same density at a given position. Conversely, $\Gamma_\nHI$ does not appear to be degenerate with other parameters. 
This means that, for observed \ha\ SB and $N_\nHI$, $\Gamma_\nHI$ is largely independent upon details on galaxy models, being primarily dependent on the SB at the ionisation front.

\begin{figure}
	\centerline{\includegraphics[width=0.5\textwidth]{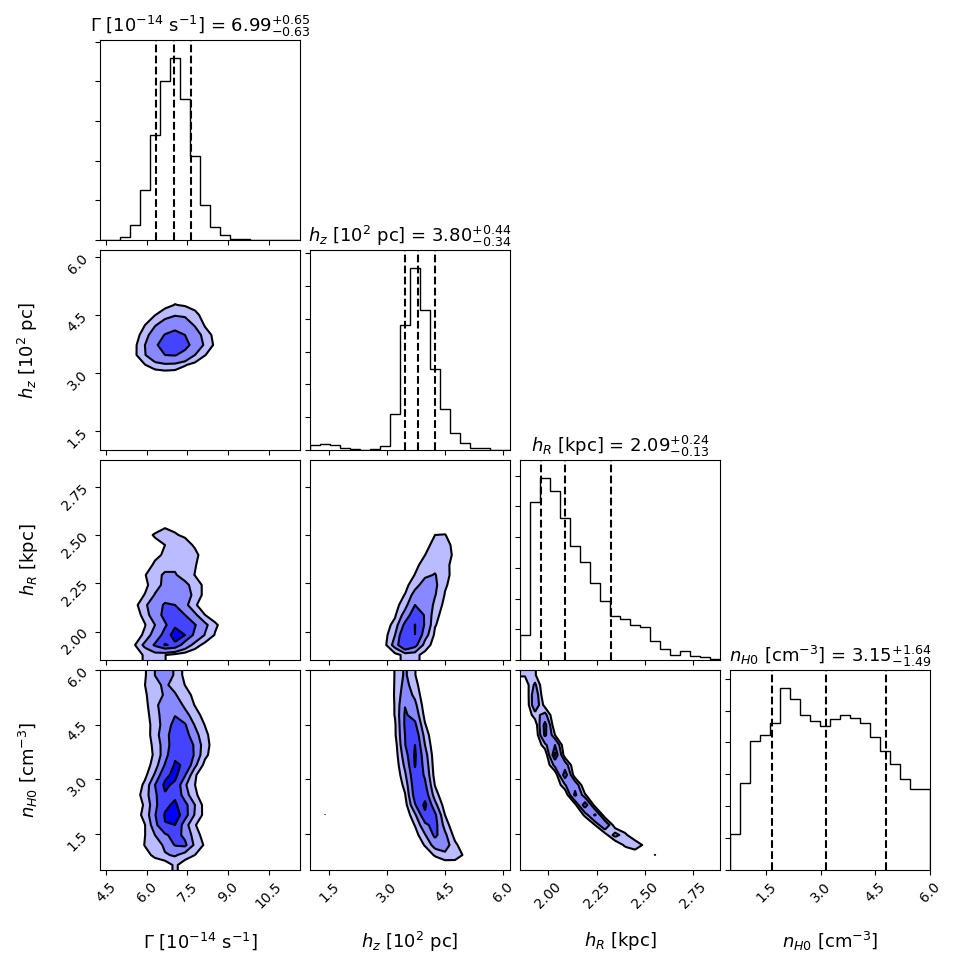}}
	\caption{Same of Fig.~\ref{fig:corner}, but after having artificially reduced the errors on the \ha\ SB by a factor of 5.}
	\label{fig:corner01}
\end{figure}

We further checked that the uncertainties on the determination of $\Gamma_\nHI$ scales approximatively with the error on the \ha\ SB. By artificially reducing the error on the SB data by a factor of 2 and 5 we obtain $\Gamma_{-14} =7.03^{+1.57}_{-1.50}$ and $6.99^{+0.65}_{-0.63}$, respectively. Errors on the other parameters are almost unchanged (see Fig.~\ref{fig:corner01} for the case in which the errors on the \ha\ SB are reduced by a factor of 5). 

We repeated the same exercise excluding any information from 21-cm data. In this case the uncertainties on the  $\Gamma_\nHI$ are much less affected by any reduction of the uncertainties on the \ha\ SB. As an example, the test with the error on SB data reduced by a factor of 5 leads to a corresponding reduction on the error of $\Gamma_\nHI$ by less than a factor of 2. This highlights the importance of joint analysis of \ha\ {\it and} 21-cm data.

%

%
%

\section{Conclusions}


Thanks to pilot MUSE observations \citep{fumagalli17}, we detected \ha\ emission in the outskirts of the nearby edge-on spiral UGC 7321, at the spatial location, wavelength position, and intensity expected for \HI in ionisation equilibrium with the cosmic extragalactic UVB.

By means of a large set of radiative transfer galaxy models, and MCMC statistical analysis, we translated our detection into a value of the UVB at $z\simeq 0$, which is consistent with several other independent measurements obtained with different techniques. When compared to empirical synthetic models of the UVB, our result strengthens the notion that the low-redshift ionising photon budget is largely dominated by AGNs, with star forming galaxies giving a marginal contribution. This is consistent with the low fraction of ionising photons leaking into the IGM generally measured in bright galaxies \citep[see, e.g.,][and references therein]{kakiichi18}. Incidentally, the decline of the AGN number density at $z\gta 2$ \citep[but see][]{giallongo15} requires that such escape fraction has to be on average much larger, $\gta 15$\%, in high-redshift galaxies if they are responsible for the reionisation of the IGM at $z\gta6$ \citep[e.g.,][]{khaire16}. It is interesting to note that, in such picture, the declining population of AGNs and the rising escape fraction from star forming galaxies must concur to produce an almost constant ionisation rate, $\Gamma_\nHI \simeq 10^{-12}$ s$^{-1}$, in the redshift range $2\lta z\lta 6$ \citep[e.g.,][]{becker13}.    
 
Further observations are required to confirm our measurements and track with new techniques the amplitude of the UVB with resdhift. Our MCMC analysis demonstrates that the derived value of $\Gamma_\nHI$ is robust against large variations in the galaxy model, and that the uncertainties are mainly driven by the errors associated with the observed \ha\ SB. This implies that future, deeper observations using a similar technique would be able to put much more stringent constraints on the value of the UVB. As we have shown, a reduction of the uncertainties on the measured \ha\ SB would lead to a proportional reduction in the $\Gamma_\nHI$ error estimate. In this respect, we anticipate that new MUSE observations of the same target (but in two different galaxy's regions) are currently scheduled at VLT (PID 101.A-0042, PI Fumagalli). These will test more robustly the origin of the detected signal and are expected to yield a more precise and accurate determination of the \HI photoionisation rate at $z\simeq 0$.

\section*{Acknowledgements}
MF acknowledges support by the Science and Technology Facilities Council (grant number ST/P000541/1). SC gratefully acknowledges support from Swiss National Science Foundation grant PP00P2\_163824. This project has received funding from the European Research Council (ERC) under the European Union's Horizon 2020 research and innovation programme (grant agreement No.757535)


\bibliographystyle{mnras}
\bibliography{../mybib}

\end{document}